\documentclass[referee]{raa} 
\usepackage{booktabs}
\usepackage{graphicx,times}
\usepackage{natbib,amssymb}
\usepackage{float}
\usepackage{ulem}
\usepackage[pagebackref=true]{hyperref}
\usepackage{subcaption}
\usepackage[utf8]{inputenc}
\usepackage{amssymb,amsmath}
\bibpunct{(}{)}{;}{a}{}{,}

\providecommand{\bm}[1]{\mbox{\boldmath$#1$\unboldmath}}

\begin{document}
	\title{Prospects of the multi-channel photometric survey telescope in the cosmological application of Type Ia supernovae}

\author{Zhenyu Wang\inst{1,2,3,4}
	 \and Jujia Zhang \inst{1,3,4}
	 \and Xinzhong Er \inst{5}
	 \and Jinming Bai \inst{1,3,4}
}

\institute{
    Yunnan Observatories, Chinese Academy of Sciences,Kunming 650216, China; 
    \it{jujia@ynao.ac.cn} \\
   \and
   University of Chinese Academy of Sciences, Beijing 100049, China; \\
   \and
   Key Laboratory for the Structure and Evolution of Celestial Objects, Chinese Academy of Sciences, Kunming 650216, China\\
   \and
   International Centre of Supernovae, Yunnan Key Laboratory, Kunming 650216, China\\
   \and
   South-Western Institute for Astronomy Research, Yunnan University, Kunming 650500, China;\\
   \vs\no
   {\small Received 20xx month day; accepted 20xx month day}}

\abstract {The Multi-channel Photometric Survey Telescope (Mephisto) is a real-time, three-color photometric system designed to capture the color evolution of stars and transients accurately. This telescope system can be crucial in cosmological distance measurements of low-redshift (low-$z$, $z$ $\lesssim 0.1$) Type Ia supernovae (SNe Ia). To optimize the capabilities of this instrument, we perform a comprehensive simulation study before its official operation is scheduled to start. By considering the impact of atmospheric extinction, weather conditions, and the lunar phase at the observing site involving the instrumental features, we simulate the light curves of SNe Ia obtained by the Mephisto. The best strategy in the case of SN Ia cosmology is to take the image at an exposure time of 130 s with a cadence of 3 days. In this condition, Mephisto can obtain hundreds of high-quality SNe Ia to achieve a distance measurement better than $4.5\%$. Given the on-time spectral classification and monitoring of the Lijiang 2.4 m Telescope at the same observatory, Mephisto, in the whole operation, can significantly enrich the well-calibrated sample of supernovae at low-$z$ and improve the calibration accuracy of high-$z$ SNe Ia. 
	\keywords{transients: supernovae; cosmology: cosmological parameters} }

\titlerunning{SNe Ia Cosmology in Mephisto}

\authorrunning{Wang et al.}

\maketitle

\section{Introduction}\label{introduction}
As the most reliable indicators of cosmological distance so far, Type Ia supernovae (SNe Ia) play a crucial role in unveiling the accelerated expansion of the universe (\citealp{1998AJ....116.1009R, 1999ApJ...517..565P}). Additionally, SNe Ia have greatly contributed to the discovery of the Hubble tension \citep{2021ApJ...912..150D}. Despite the proposal of dark energy as an explanation for the accelerated expansion of the universe, our understanding of its fundamental properties remains limited. As the most straightforward observational tool for constraining the parameters associated with dark energy, SNe Ia achieve a high level of accuracy when employed for distance measurements (\citealp{2013PhR...530...87W}). However, the recent accuracy of SNe Ia (e.g., $\lesssim 7\%$, \citealp{2019ApJ...872L..30A}) falls short in constraining cosmological parameters and determining the nature of dark energy.

The crucial factor in constraining the range of cosmological parameter values lies in accurately ascertaining the distances of SNe Ia at high-$z$ ( $z>$ 0.5). The application of SNe Ia in cosmological distance measurement depends on some empirical relations between the absolute peak brightness and shape of light curves, for example, the width-luminosity relation (WLR, usually known as the Phillips relation; \citealp{1993ApJ...413L.105P}). The original form of the Phillips relation is  
\begin{equation}
	M(\lambda)_{max}=a+b\Delta m_{15}(\lambda),
\end{equation}
where $M(\lambda)_{max}$ represents the maximum absolute magnitude at a particular band (usually using the $B$-band), and $\Delta m_{15}$ is the decline rate from the peak luminosity to 15 days later. The coefficients $a$ and $b$ are measured by dozens of well-observed low-$z$ SNe Ia.
Comparing with the low-$z$ sample, one can derive the intrinsic brightness of SNe Ia at a corresponding band for a given redshift and thus measure its luminosity distance. For example, at $z$ = 0.01, the rest-frame 4310\AA\, flux is observed in the $B$-band, while at $z$ = 1.01 this same flux is roughly observed in the $I$-band. To accurately determine the distances of SNe Ia at various redshifts, it is imperative to possess knowledge regarding the wavelength-dependent evolution of WLR. This calibration predicament in SN Ia cosmology is the primary source of systematic error in constructing the SNe Ia Hubble diagram. Consequently, enhancing calibration precision at low-$z$ becomes indispensable for refining measurements of cosmological parameters using SNe Ia. Moreover, low-$z$ supernova samples not only help to reduce the uncertainties associated with high-$z$ SN Ia measurements but also provide constraints to the Hubble constant (\citealp{2009ApJ...699..539R, 2011ApJ...730..119R, 2016ApJ...826...56R, 2022ApJ...934L...7R}), which can alleviate the Hubble tension.

For low-$z$ samples, the current largest source of systematic error is the calibration. The existing low-$z$ SNe Ia come from different telescopes, with some using Bessel-like $UBVRI$ filters and some utilizing SDSS-like $ugriz$ filters. Differing filter systems present a significant challenge for calibration at the level of a few percent. Even with the same filter system, subtle differences can exist in the transmissions. The most direct approach to reducing this error is to obtain additional samples using a consistent telescope system. Extinction is another factor of systematic uncertainty. For the purpose of calculating extinction, accurate color measurement is essential. Additionally, variations in ignition positions (central versus off-center) and explosion mechanisms (deflagration versus detonation) contribute to the diversity of SN Ia observations.

Furthermore, WLR is not only a function of wavelength but also depends on the spectral parameters of SNe Ia. Besides the outliers that cannot be standardized by any kind of WLR, the normal SNe Ia can be further divided into some subclasses depending on the spectral features around the peak brightness, e.g., the equivalent width (EW, \citealp{2006PASP..118..560B}), velocity (\citealp{2009ApJ...699L.139W}), and temporal velocity gradient (\citealp{2005ApJ...623.1011B}) of Si II $\lambda $6355. The fine classification is an efficient way to improve the accuracy of SNe Ia in distance measurement. For example, \citet{2009ApJ...699L.139W} divided the normal SNe Ia into two subclasses depending on the velocity of Si II 6355 around the maximum brightness and improved the accuracy of the Hubble diagram from 0.178 mag to 0.125 mag. \citet{2012AJ....143..126B} use different coefficients to fit each subclasses of various SNe Ia depending on the EW of Si II lines. Thus, to create a consistent sample, it is crucial to have three spectra for every well-observed SN Ia that covers the spectral evolution from a few days before to a few days after the $B$-band maximum.

In this paper, we investigate the possibility of the Multi-channel Photometric Survey Telescope (Mephisto, \citealp{2020SPIE11445E..7MY}) for low-$z$ calibration of SNe Ia. Mephisto is a 1.6-meter wide-field high-precision telescope with a 2 deg$^2$ field of view. It is led by Yunnan University and co-built by Yunnan Observatories and Nanjing Institute of Astronomical Optics \& Technology. This telescope can simultaneously image the sky in three bands ($ugi$ or $vrz$) and obtain real-time color information. It boasts high color calibration accuracy, which can effectively reduce errors introduced by extinction and reddening. Moreover, the available samples around $z=0.1$ so far are low (\citealp{2019ApJ...872L..30A}). As a 1.6 m-aperture wide-field survey telescope, it has the ability to obtain a great number of high-quality samples.
Based on multi-color photometry, it can trigger the Lijiang 2.4 m Telescope (LJT,  \citealp{2015RAA....15..918F}) to perform follow-up spectral observation. We can get the spectral evolution to help us identify the subclass. Hence, we can remove SNe Ia with larger deviations in the analysis, and thus reduce the uncertainty in the Hubble diagram.

To improve the observation efficiency of Mephisto, we conduct observation simulation and cosmological fitting to produce an appropriate observation strategy. We use SNCosmo (\citealp{barbary_2022_7117347}) in this study to simulate the light curves of SNe Ia for Mephisto. Section~\ref{Mephisto and Lijiang Observatory} provides a concise introduction to the fundamental information of Mephitso and the Lijiang Observatory. Section~\ref{The simulation of SN Ia} delves into the simulation of SN Ia light curves and the photometric accuracy under different exposure durations. The cosmological parameter estimation method is  detailed in Section~\ref{Calibration and Constraints on Cosmological Parameters}. Section~\ref{result} is the presentation of the results, and there we have provided appropriate observation strategies. Section~\ref{Discussion} presents the results of the estimation and provides suggestions for future observations.  

\section{Mephisto and Lijiang Observatory}
\label{Mephisto and Lijiang Observatory}
Mephisto innovatively uses a Ritchey-Chrétien (RC) system with correctors and film-coated cubic prisms, so that the focal plane of the telescope is under the main mirror and there is a large distance between the focal plane and the main mirror. Therefore, the color separation system and multiple cameras can be placed in the rear optical path while ensuring a high-quality image in the entire field of view. The telescope is equipped with three high-quality CCD cameras, which can capture real-time color images of the same patch of sky in three different bands ($ugi$ or $vrz$), providing valuable color information.

\begin{figure}
	\centering
	\includegraphics[scale=0.92]{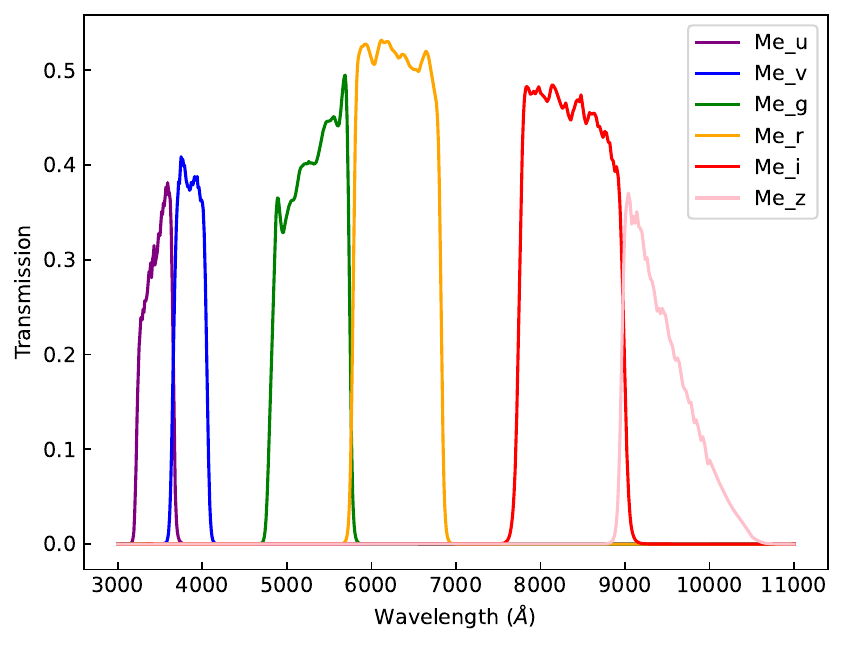}
	\caption{The current total efficiency of Mephisto in $u, v, g, r, i$ and $z$ bands. \label{fig1}}
\end{figure}

The current total efficiency of Mephisto is depicted in Figure~\ref{fig1}. The filters used by Mephisto are not the most popular system among observers, e.g., $UBVRI$ of Johnson-Cousins or $ugriz$ of SDSS. The specific choice of these filters depends largely on their scientific objectives. Mephisto's survey mode includes both W-survey and D/H/M-survey modes. The W mode focuses on 2*20 second exposures of observable areas in the northern hemisphere ($\delta>$15°, ~26,000 square degrees). The D/H/M models, on the other hand, sample at daily, hourly, and minute cadence, respectively, for time-domain astronomy observations. The observation strategy adopted in this paper is based on D mode.

Mephisto is situated at the Lijiang Observatory administered by Yunnan Observatories, located in Gaomeigu, Lijiang City, in Yunnan Province, China, at an altitude of 3200 meters. The site benefits from an effective observation time of more than 2000 hours per year. The average seeing is about 1\arcsec.17, with 25$\%$ of the time experiencing seeing below 0\arcsec.91 and 75$\%$ of the time experiencing seeing below 1\arcsec.33 (\citealp{2020RAA....20..149X}).
Overall, the Lijiang Observatory is an outstanding site for optical surveys.  

\section{The simulation of SNe Ia}
\label{The simulation of SN Ia}
The simulation of SN Ia light curves is discussed in Section~\ref{light curve}. We take atmospheric extinction, seeing, telescope efficiency, weather, and lunar phase into consideration. We explore the impact of exposure time on the accuracy of light measurement and distance measurement in Section~\ref{exposure time}. In Section~\ref{Sepctral}, we list some well-observed spectra captured by LJT. 

\subsection{SN Ia light curve}
\label{light curve}
The modified spectral adaptive light curve template (SALT2,  \citealp{2007A&A...466...11G}) model produces a spectral energy distribution (SED) for the following simulation. 
The flux is defined as follows
\begin{equation}
	F(t,\lambda)=x_0\times [M_0(t,\lambda) + x_1M_1(t,\lambda)]\times exp[c \times CL(\lambda)],
\end{equation}
where $\lambda$ and t represent the rest-frame wavelength and time respectively, $M_0$($t$, $\lambda$) and $M_1$(t, $\lambda$) describe the temporal variation of SED, CL($\lambda$) is the average color correction law, $x_0$, $x_1$ and $c$ are the fitting parameters of the light curves for SNe Ia, $x_0$ is the flux normalization parameter, $x_1$ is the stretch parameter, and $c$ is color.
In addition to atmospheric extinction, seeing, and telescope efficiency, we consider several other factors, including:

\begin{figure}[H]
	\centering
	\includegraphics[scale=0.94]{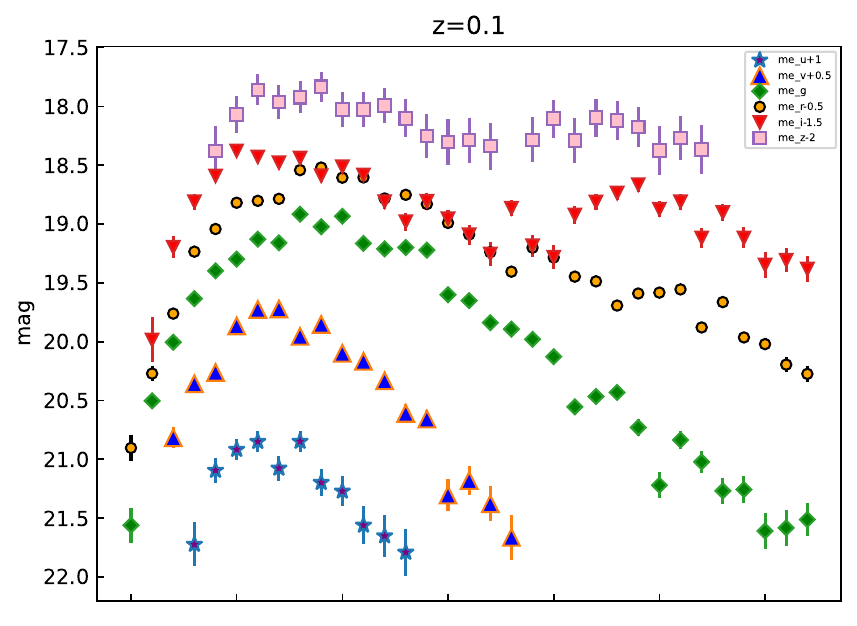}	
	\includegraphics[scale=0.94]{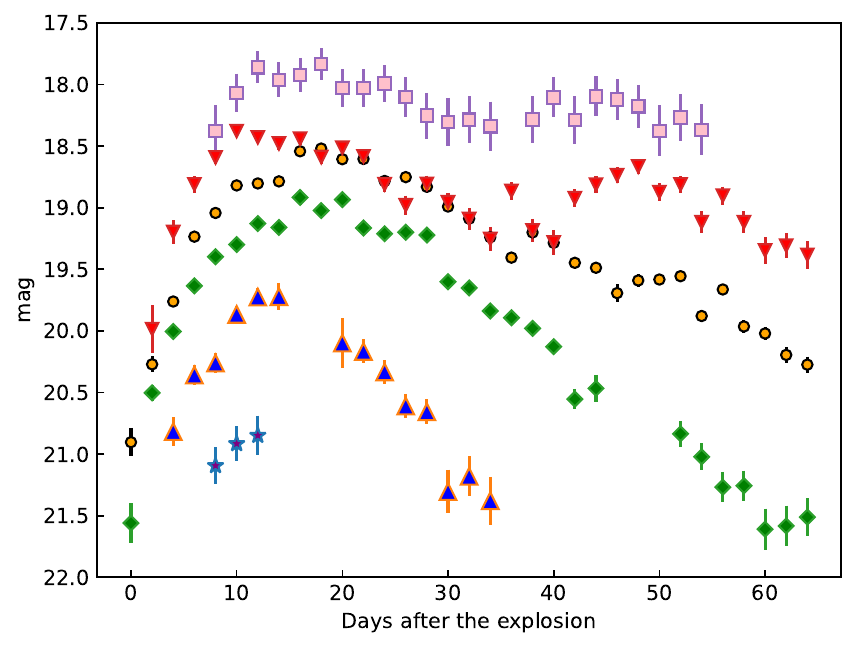}
	
	\caption{This figure simulates an SN Ia with $z$ = 0.1, taking atmospheric extinction, seeing, weather, and lunar phase into account. The exposure time is 130 s. The absolute magnitude in the $B$-band is -19.3 mag. The top panel shows the observation data without the influence of the lunar phase, while the bottom panel includes the lunar phase.  \label{fig2}}
\end{figure}

\textbf{Weather: }Lijiang experiences a notable rainy season from June to September. The number of nights suitable for observation during this period is relatively limited. This paper primarily focuses on analyzing the weather patterns during the dry season. Based on the statistics of the weather conditions (\citealp{2020RAA....20..149X}), usually there are about 200 nights in one year which are suitable for observations. In this study, we adopt 200 observational nights per year in the simulation. The occurrence of supernovae is uniformly and randomly distributed among these 200 days. Furthermore, weather conditions are also taken into account. Even during the dry season, there may be occasional instances where observation is hindered due to factors such as precipitation or high humidity. Therefore, a few data points are randomly omitted from the observed dataset to account for these circumstances.

\textbf{Lunar phase:} The influence of the lunar phase on observation is predominantly concentrated in the $u$ and $g$ bands, which correspond to the $u$, $v$, and $g$ bands of Mephisto. We incorporated the effect of lunar phase into these three bands, based on data from ref. \citet{2020RAA....20..149X}, as per the corresponding lunar calendar during the time of observation. The observation data with and without the influence of the lunar phase are presented in Figure~\ref{fig2}. As we can see, the influence of the lunar phase decreases the number of observable days in the $u$ and $v$ bands.

\textbf{data:} We generate a batch of SNe Ia with the following conditions. In our simulation, the redshifts are less than 0.5. The volumetric rate of SNe Ia we adopt in this simulation is $1.0 \times 10^{-4}$ yr$^{-1}$ Mpc$^{-3}$ according to ref. \citet{Rodney_2014}. The redshift distribution of 4957 SNe Ia with 200 days of observation time and a 180 $deg^2$ sky area is shown in the upper panel of figure \ref{fig4}. The absolute magnitude follows a normal distribution with a mean of -19.26 and a standard deviation of 0.2 (\citealp{2014AJ....147..118R}). The distributions of the parameters $x_1$ and $c$ obey the following functions (\citealp{2016ApJ...822L..35S}):
\begin{equation}
	P(x)=\left\{ 
	\begin{aligned}
		Ae^{(-(x-\overline{x})^2)/2\sigma_1^2},  x<\overline{x}; \\
		Ae^{(-(x-\overline{x})^2)/2\sigma_2^2},  x\geq\overline{x}. \\
	\end{aligned}
	\right.
\end{equation}
where $A=\sqrt{2/\pi}(\sigma_1+\sigma_2)^{-1}$, and $\overline{x}$ = 1.142, $\sigma_1$ = 1.652, and $\sigma_2$ = 0.104 for $x_1$, and $\overline{x}$ = -0.061, $\sigma_1$ = 0.023, $\sigma_2$=0.083 for c (\citealp{2023SCPMA..6629511L}). With these distributions, we have compiled a comprehensive library of SN Ia light curves, which serves as a valuable resource for further analysis.

\subsection{exposure time}
\label{exposure time}
\begin{figure}
	\centering
	\includegraphics[scale=0.94]{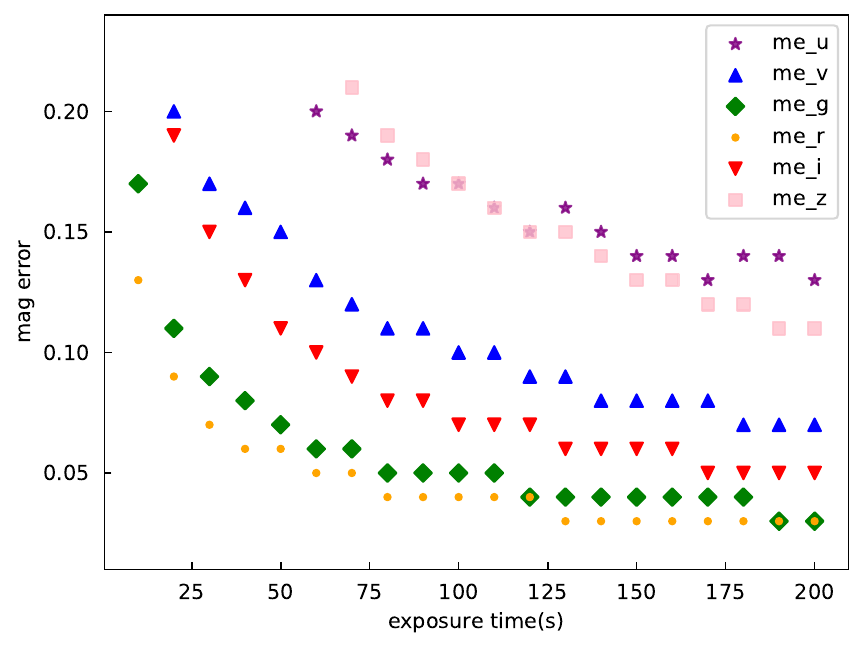}
	
	\caption{ The horizontal axis is exposure time, and the vertical axis is the average mag error over the 10 days before and after the $B$-band maximum. The exposure time is from 10 to 200 s, with two decimal places retained for the result. We select an SN Ia with a redshift of 0.1. \label{fig3}}
\end{figure}

\begin{table}
	\centering
	\caption{The fitting results under different exposure times, and $\mu$ represents the distance modulus. \label{tab1}}
	\setlength{\tabcolsep}{27mm}{
		\begin{tabular}{ cc }
			\toprule
			exposure time	& $\mu$ \\
			\midrule
			20 s			    & $39.02\pm{0.21}$ \\
			60 s	 	        & $38.84\pm{0.11}$ \\
			80 s				& $38.83\pm{0.10}$ \\
			130 s				& $38.82\pm{0.09}$ \\
			\midrule
			\bottomrule
	\end{tabular}}
\end{table}
We test the photometric accuracy at different exposure times. The results are plotted in Figure~\ref{fig3}. Due to the sensitivity of the SALT2 fitting to observation points near the maximum brightness, we select a total of 10 data points before and after the maximum and evaluate the quality of photometric measurements based on their errors. Additionally, for bands with fewer than 10 data points, we include all available data in our analysis.  

As expected, the photometric error generally decreases exponentially with increasing exposure time. However, lower apparent magnitudes become visible in the u-band with increasing exposure time, which leads to an overall increase in the average photometric error in the $u$-band. Only data from the $g$ and $r$ bands can be captured when the exposure time is limited to 10 s. Data of $v, g, r$, and $i$ bands are available in all four bands when the exposure time is longer than 20 s. Data from the $u$ band also becomes observable when the exposure time is further increased to 60 s. Data from all six bands can be detected when the exposure time exceeds 70 seconds. An increase of 10 s in exposure time does not bring significant improvements in accuracy at 80 and 130 s. We take into account the SALT2 fitting accuracy under various exposure times for further analysis. Table~\ref{tab1} displays the results of the fitting. Fitting results show very small and close errors when the exposure time exceeds 60 s. This indicates that the telescope can achieve good observation accuracy for SNe Ia with redshifts near 0.1. The error in the distance modulus decreases to $~0.1$ mag when the exposure time exceeds 80 s. This means the distance uncertainty is $\lesssim 4.5\%$ according to the law of propagation of uncertainties. We take 80 and 130 s into consideration for further exploration. 

\subsection{Spectral diversity of well-observed SNe Ia}
\label{Sepctral}

Spectral follow-up of LJT is crucial for achieving a homogeneous SN Ia sample from Mephisto. For example, including three spectra at $t\sim -7, 0, +7$ days post-maximum is vital in obtaining essential information for diversity studies, encompassing the examination of EW, velocity, and velocity graduation. Consequently, we can effectively exclude subclass samples exhibiting more significant dispersion, thereby reducing distance measurement errors arising from scatter and enhancing the fitting accuracy of cosmological parameters.

\begin{table}
	\centering
	\caption{The observation results of SNe Ia with a redshift of 0.1 under different cadences.\label{SN Ia}}
	\setlength{\tabcolsep}{5mm}{
		\begin{tabular}{ cccccc }
			\toprule
			SN Ia		&$\Delta m_{15}(B)$  & $M_B$ &type(wang) &type(branch) & ref. \\
			\midrule
			13dy	 & 0.90      & -19.65 & NV & CN &  \citealp{2016AJ....151..125Z}\\
			
			09ig     & 0.90      & -19.46 & HV & $\cdots$  & \citealp{2013ApJ...777...40M} \\
			
			12fr	 & 0.85		& -19.49 & HV & SS  & \citealp{2014AJ....148....1Z} \\
			
			15bq	 & 0.82		& -19.68 & $\cdots$ & SS  & \citealp{2022ApJ...924...35L} \\
			
			19ein	 & 1.35		& -18.71 & HV & BL & \citealp{2022MNRAS.517.4098X}  \\
			
			11fe	 & 1.18		& -19.40 & NV  & CN  & \citealp{2016ApJ...820...67Z} \\
			\midrule
			\bottomrule
	\end{tabular}}
\end{table}

\begin{figure}
	\centering
	\includegraphics[scale=1.1]{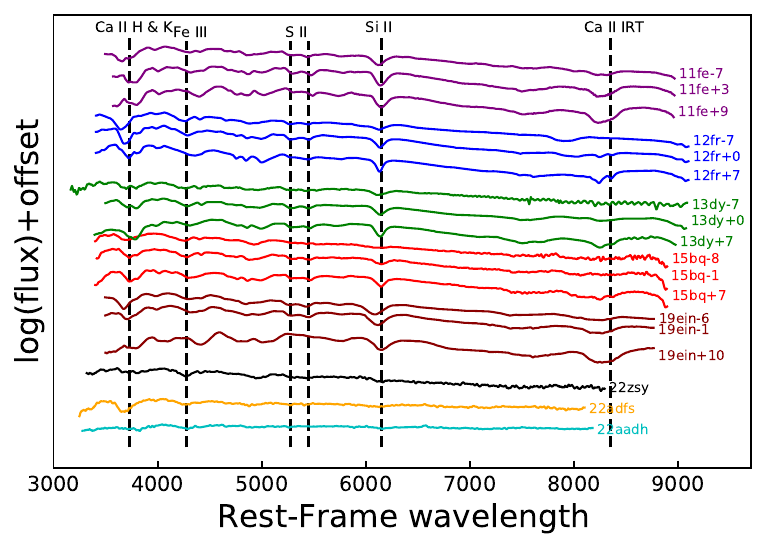}
	
	\caption{Temporal spectra of different subclasses of SNe Ia at a few days before and after the $B$-band maximum. The spectrum of 13dy-7 is from ref~\citet{2013ApJ...778L..15Z}, and the others are from LJT.  \label{spec}}
\end{figure}

We present the photometric data of SNe Ia in Table~\ref{SN Ia}, while Figure~\ref{spec} shows spectra of selected well-observed SNe Ia. Notably, supernovae with similar $\Delta m_{15}$ values may exhibit varying absolute magnitudes, e.g., SN 2013dy and SN 2009ig. 
Through the measurement of the velocity of SiII $\lambda$6355 SN 2013dy is classified into the normal-velocity (NV) type, while SN 2009ig is classified into the high-velocity (HV) type in the Wang diagram (\citealp{2009ApJ...699L.139W}). The $\Delta m_{15}$ and $M_B$ of SNe 2012fr and 2015bq are similar. They are both divided into shallow silicon (SS) due to the small EWs ($\le 60\AA$) of Si II $\lambda$6355 (\citealp{2006PASP..118..560B}). However, SN 2012fr shows narrower Si II $\lambda$6355 and Ca II IRT profiles, with generally deeper absorption lines than the 91T-like events. In the early spectra, there are strong high-velocity features of Si II $\lambda$6355 that are not seen in other 91T-like events \citep{2014AJ....148....1Z}. SN 2019ein is a 91bg-like supernova. Its peak brightness is significantly lower than normal SNe Ia. During maximum, the 91bg-like events show weak or absent Fe II lines and strong absorption lines of intermediate elements. These supernovae do not follow the WLR. SN 2011fe is a typical normal SN Ia. We also list some SNe Ia with $z > 0.05$, e.g., SN 2022zsy ($z \sim 0.06$), SN 2022aadh ($z \sim 0.076$), and SN 2022adfs ($z \sim 0.088$). 

The collaboration between LJT and Mephisto enables us to acquire a substantial quantity of samples featuring spectra at $z\sim0.1$ for subsequent classification based on multidimensional spectral information. This information facilitates the elimination of subclasses that significantly deviate from the WLR, resulting in a more homogeneous sample set. Additionally, distinct subclasses of supernovae exhibit varying Phillip relationship coefficients, effectively mitigating the dispersion observed in the SNe Ia Hubble diagram.

\section{Calibration and Constraints on Cosmological Parameters}
\label{Calibration and Constraints on Cosmological Parameters}

To achieve higher accuracy, we conduct a preliminary screening on the data before performing the light curve fitting. We remove some supernovae with a limited number of observed data points. The filtering criteria are as follows: 
\begin{enumerate}
	\item[1)] signal-to-noise ratio (S/N) greater than 5; 
	\item[2)] The sum of all the bands is at least 15 data points;  
	\item[3)] Data are available in at least three different bands; 
	\item[4)] After applying the aforementioned selection criteria, we retained only the supernovae with successful convergence in the SALT2 fitting. 
\end{enumerate}
We refer to SNe Ia that have passed the screening as good-samples. Table~\ref{tab2} presents the number of supernovae obtained under the observation conditions of a 200 day time span and a 180 deg$^2$ sky area. Mephisto has many scientific goals, such as the search for extremely metal-poor stars and transient sources. In our simulation, we focus on a 180 $deg^2$ sky area for our SNe Ia survey, which accounts for about 20\% of the observing time. Our analysis primarily focuses on the cases with cadence of 2 days, 3 days, or 5 days, considering exposure times of 80 s and 130 s respectively.

\begin{table} [H]
	\centering
	\caption{The number of SNe Ia that can be observed after screening at different cadence and exposure times. The observation time is one year. \label{tab2}}
	\setlength{\tabcolsep}{17mm}{
		\begin{tabular}{ccc}
			\toprule
			\textbf{Cadence}	& \textbf{exposure time}	& \textbf{Number}\\
			\midrule
			2 d		& 130 s   & 1094\\
			2 d		& 80 s  & 762\\
			3 d		& 130 s	& 979\\
			3 d		& 80 s  & 616\\
			5 d     & 130 s   & 566\\
			5 d     & 80 s  & 312\\
			\midrule
			\bottomrule
	\end{tabular}}
\end{table}
Various factors can influence the details of an SN Ia explosion, including the precise location of ignition, whether it is at the center or off-center, and the dynamical burning mechanism (e.g., subsonic deflagration or supersonic detonation). Additionally, the circumstellar environment in which the explosion occurs plays a significant role. These factors collectively contribute to the observational diversities and result in distinct light curve parameters. The dispersion caused by these influential factors can substantially impact accurately fitting cosmological parameters. Therefore, prior data calibration becomes imperative before undertaking cosmological parameter fitting procedures. Our study employed the minimizing $\chi^2$ methodology  \citep{2011ApJ...740...72M} to produce the best fitting result.
\begin{equation}
	\chi^2(\alpha,\beta)=\sum\limits_{i=1}[m_{xi}-{\mu}(z_i)+{\alpha}x_{1i}-{\beta}c_i-M_{bin}(z)]^2/({\sigma}_i^2+{\sigma}_{int}^2),
\end{equation} 
where $m_{xi}$, $x_{1i}$, and $c_i$ are the best-fit parameters of the SALT2 model for the $i-th$ SN Ia. $M_{bin}$ is a constant in each redshift bin. The number of bins exceeds 30, with each redshift bin having a width of 0.01. The definition of $M_{bin}$ is as follows 
\begin{equation}
	M_{bin}(z_b)=m_{x}-{\mu}(z)+{\alpha}x_{1}-{\beta}c,
\end{equation} 
where $m_x=-2.5logx_0$, $\mu(z)$ is the distance modulus, and the redshift $z_b$ represents the central point of each bin. One of the key benefits of this approach is its robustness against variations in wavelength bands. The difference between $m_x$ and $m_0$ is about 10 mag. While a cosmological model is employed to calculate $\mu(z)$ in this process, it has been shown in ref.  \citet{2011ApJ...740...72M} that the values of $\alpha$ and $\beta$ are not affected by the choice of cosmological parameters when a large number of bins are used.  
The definition of $\sigma_{int}$ is as follows 
\begin{equation}
	{\sigma}_{int}^2=V_{m_x}+\alpha^{2} V_{x_1}+ \beta^2 V_c+2\alpha V_{m_x, x_1} - 2\beta V_{m_x,c} - 2\alpha\beta V_{x_1,c} ,
\end{equation} 
where $V$ represents the intrinsic covariance matrix of parameters, $V_{m_x}$, $V_{x_1}$, and $V_c$ are the diagonal elements of the covariance matrix, $V_{m_x, x_1}$, $V_{m_x,c}$, and $V_{x_1,c}$ are the corresponding off-diagonal elements. $\sigma_{i}^2$ is defined as follows 
\begin{equation}
	{\sigma}_{i}^2=V_{m_{xi}}+\alpha^{2} V_{x_{1i}}+ \beta^2 V_{c_i}+2\alpha V_{m_{xi}, x_{1i}} - 2\beta V_{m_{xi},c_i} - 2\alpha\beta V_{x_{1i},c_i} ,
\end{equation} 
Similar to the definition of $\sigma_{int}^2$, the covariance matrix consists of $m_x$, $x_1$, and $x_0$ for the $i-th$ SN Ia.  
Restrictions on cosmological parameters can be made based on the observed distance moduli $\mu$ and the theoretically calculated values of $\mu_{th}$. We consider the flat $\Lambda CDM$ model here.
We used the MCMC (\citealp{2013PASP..125..306F}) method for parameter fitting, with particular focus on the values of $H_0$ and $\Omega_{M}$. The likelihood function is defined as follows 
\begin{equation}
	lnP=-\frac{1}{2}\sum_{i=0}^n \frac{(\mu_i - \mu_{th}(z_i, H_0, \Omega_M))^2}{\sigma_{\mu_i}^2}
\end{equation}
where $\mu _i = m_x - M_0(z_b) + \alpha x_{1_i} - \beta c{_i}$,  and $\mu_{th}$ is defined as follows 
\begin{equation}
	\mu_{th}(z_i, H_0, \Omega_M)=5log(d_L/Mpc)+25
\end{equation} 
Considering a flat universe and discarding the radiation term, the luminosity distance $d_L$ is defined by 
\begin{equation}
	d_L=\frac{c(1+z)}{H_0}\int_0^z{\frac{dz'}{\sqrt{\Omega_{M}(1+z')^3+(1-\Omega_M)}}}
\end{equation} 
The absolute peak magnitude of SNe Ia in our simulation follows a normal distribution with a mean of -19.26 and a deviation of 0.2 \citep{2014AJ....147..118R}. This can lead to the production of supernovae that are very bright or very dim. We removed data that significantly deviated from the Hubble flow.

\section{Result}
\label{result}
\begin{figure}
	\centering
	\includegraphics[scale=0.8]{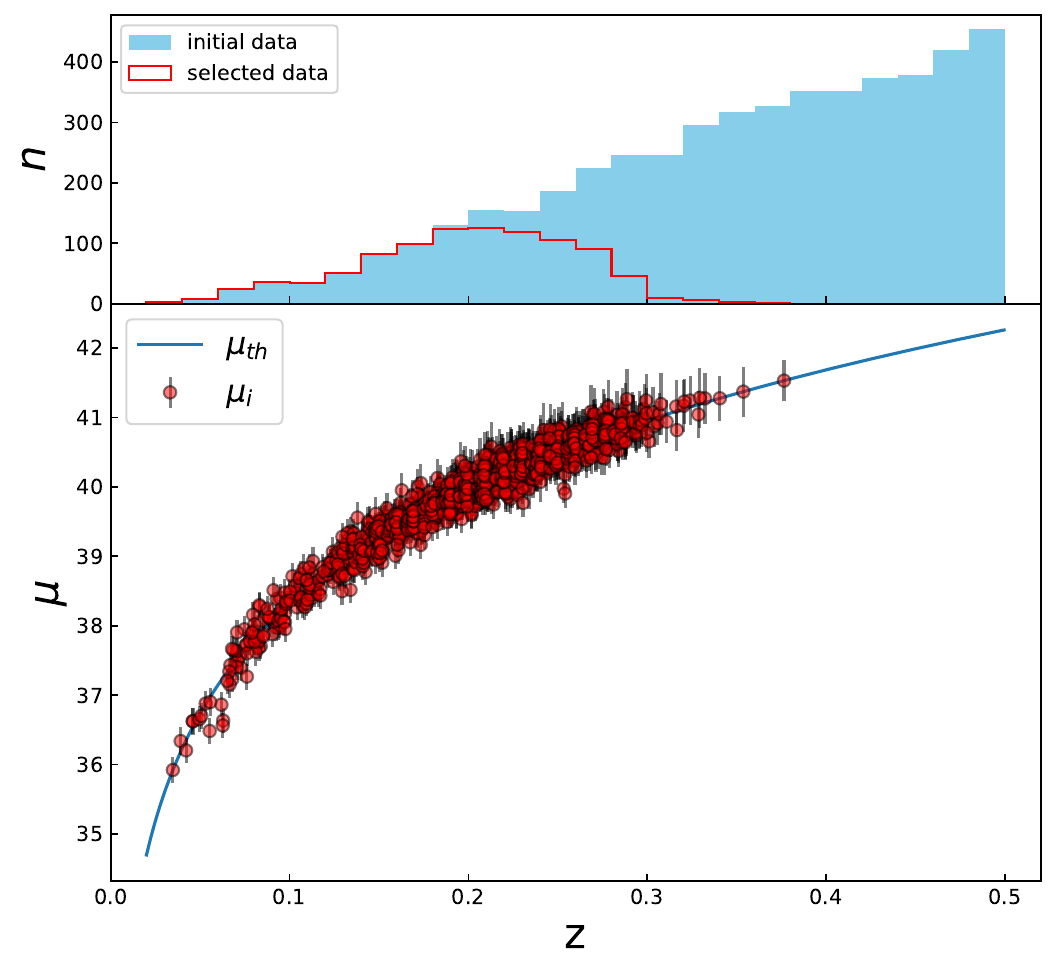}
	\caption{The top panel shows the original number-redshift distribution of simulated SNe Ia (initial data, blue histogram) and selected SNe Ia (selected data, red step line). The bottom panel displays the moduli-redshift distribution of selected SNe Ia (red points). The solid line in the bottom panel represents the theoretical prediction. The figure illustrates a scenario where the cadence is 3 days and the exposure time is 130 s for red points. The observation time is one year for both initial data and selected data. \label{fig4}}
\end{figure} 

Figure~\ref{fig4} displays the redshift distribution of the screened data. The observation strategy utilized a cadence of 2 days and an exposure time of 130 s. The top panel displays the distribution of the number of SNe Ia with redshifts. The blue histogram represents the original redshift distribution of simulated SNe Ia, while the red step line depicts the number distribution of SNe Ia after the selection process. The bottom panel displays the distribution of distance moduli with redshift. The solid line represents a flat $\Lambda CDM$ model with $\Omega _{M_0}$=0.3, $\Omega _{\Lambda _0}$=0.7, and $H_0$=70 km s$^{-1}$Mpc$^{-1}$, which are used to generate the SNe Ia data. The red points in the chart represent the distribution of SNe Ia after the selection process. Under this observational strategy, the telescope's observational limit extends slightly beyond a redshift of 0.3. For supernovae with a redshift around 0.1, there is not only good observational accuracy but also no significant selection bias. However, selection bias is apparent when the redshift exceeds 0.26.

\begin{table}
	\centering
	\caption{The observation results of an SN Ia with a redshift of 0.1 under different cadences.\label{tab3}}
	\setlength{\tabcolsep}{15mm}{
		\begin{tabular}{ ccc }
			\toprule
			Observation Strategy		&$H_0$  & $\Omega_M$\\
			\midrule
			cadence=2 days, t=130 s	 &$68.76^{+0.70}_{-0.72}$               & $0.44^{+0.07}_{-0.07}$ \\
			
			cadence=2 days, t=80 s	 &$68.83^{+0.91}_{-0.95}$				&$0.45^{+0.12}_{-0.10}$ \\
			
			cadence=3 days, t=130 s	 &$69.73^{+0.76}_{-0.77}$			    &$0.36^{+0.08}_{-0.08}$ \\
			
			cadence=3 days, t=80 s	 &$69.70^{+1.01}_{-1.05}$				&$0.37^{+0.13}_{-0.12}$ \\
			
			cadence=5 days, t=130 s	 &$70.17^{+1.00}_{-1.03}$				&$0.31^{+0.11}_{-0.11}$ \\
			
			cadence=5 days, t=80 s	 &$70.58^{+1.36}_{-1.40}$				&$0.26^{+0.19}_{-0.17}$ \\
			\midrule
			\bottomrule
	\end{tabular}}
\end{table}

\begin{figure}
	\centering
	\includegraphics[scale=0.48]{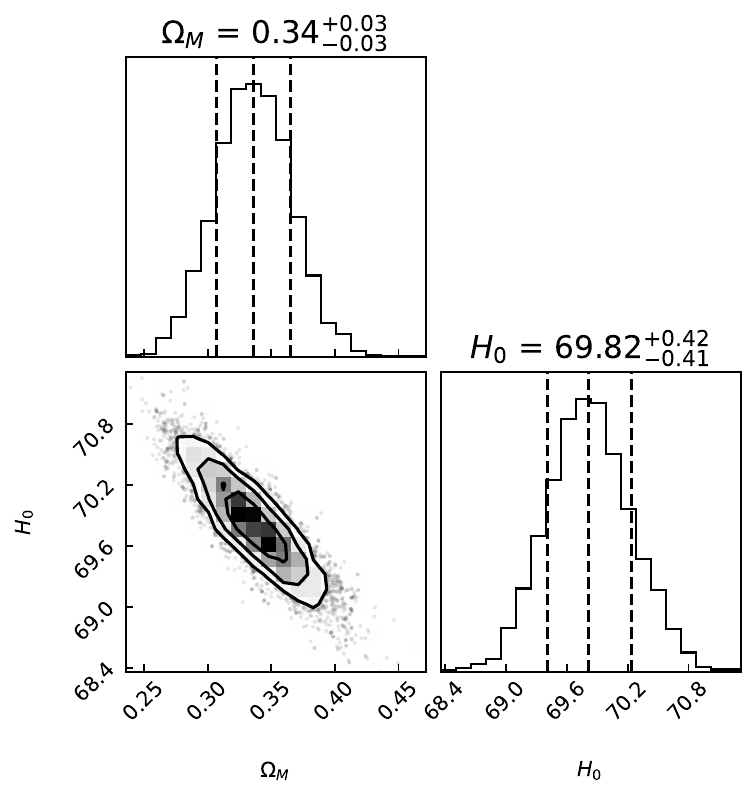}
	\hspace{0.5in}
	\includegraphics[scale=0.48]{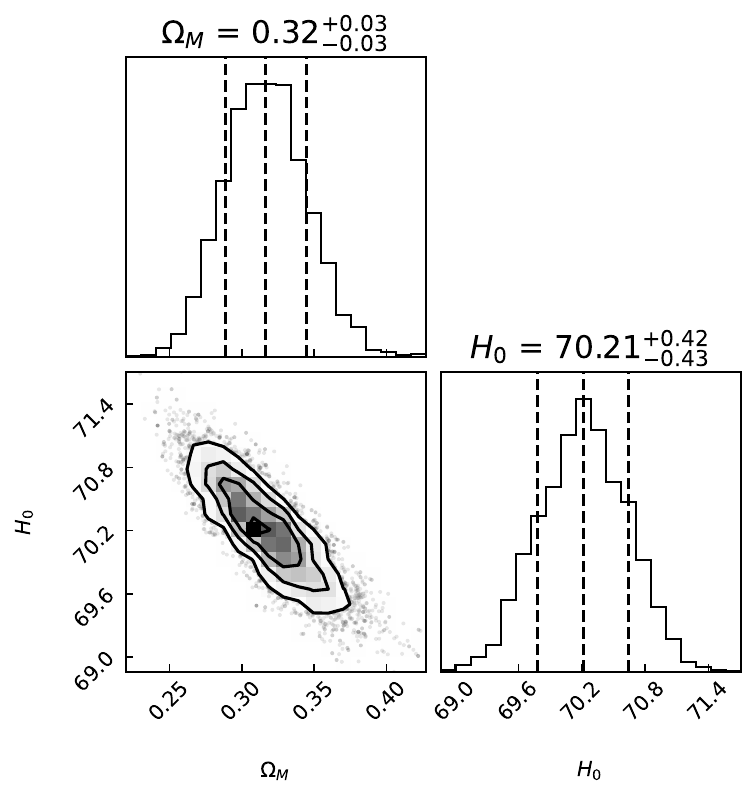}

	\caption{The MCMC fitting results include high-redshift SNe Ia. The left panel is cadence = 2 days, and the right panel is cadence = 3 days. The exposure times are both 130 s with a year of observation time.\label{fig5}}
\end{figure}
The fitting results are shown in Table~\ref{tab3}. The obtained fit results do not align with the parameters used to generate the SNe data, which will be discussed in the next section. Here we focus on the errors rather than the fit values. From the results, it can be seen that the fit values have the smallest errors for a cadence of 2 days and an exposure time of 130 s. However, compared to the strategy with a cadence of 3 days and the same exposure time, it increases the observation time by 33$\%$, resulting in only a 12.5$\%$ decrease in the error of $\Omega_M$ and almost no change in the error of $H_0$. We include 88 high-$z$ samples at redshifts higher than 0.5 (\citealp{2007ApJ...659...98R,2019ApJ...874..150B}). The improvement in precision at this level will be washed out when incorporating existing data that includes high-$z$ samples (see Figure~\ref{fig5}). Therefore, taking everything into consideration, we recommend adopting the strategy of a cadence of 3 days and an exposure time of 130 seconds for the observations.

\section{Discussion and conclusion}
\label{Discussion}

\begin{figure}
	\centering
	\includegraphics[scale=0.92]{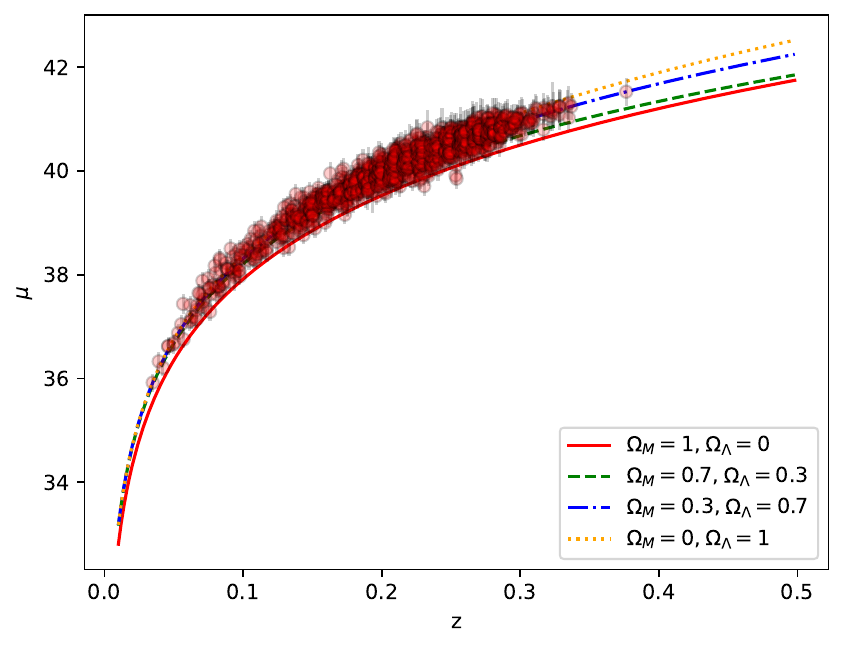}
	
	\caption{ Hubble diagram for different values of cosmological parameters. Red points represent simulated SN Ia data, similar to the data in Figure \ref{fig4}.  \label{fig6}}
\end{figure}

The results obtained by different strategies indicate a deviation from the cosmological parameters in generating the supernovae. The discrepancy is attributed to multiple factors. Firstly, a sample collected at low-$z$ provides limited constraints on cosmological parameters. Figure~\ref{fig6} illustrates the Hubble diagram at lower redshifts for different values of $\Omega_M$ and $\Omega_{\Lambda}$. It can be seen that at low redshifts, the curves predicted with different $\Omega_M$ and $\Omega_{\Lambda}$ almost overlap, indicating that the constraints on $\Omega_M$ and $\Omega_{\Lambda}$ are weak by the low-$z$ samples.  Even a slight deviation can result in significant discrepancies in the fitting values, highlighting the sensitivity of the process. Another factor to consider is the diversity exhibited by SNe Ia, which stems from variations in ignition locations, explosion mechanisms, and environments. This study does not incorporate spectral simulations, thus making it difficult to ascertain the abnormal supernovae within the simulated data. This lack of comprehensive classification of SNe Ia may introduce biases into the fitting of cosmological parameters. Moreover, it is essential to acknowledge that Mephisto filters are non-standard and may introduce certain biases during fitting procedures. Lastly, while the light curves of SNe Ia align with actual observations, any deviations between fitted and actual values can be attributed to potential limitations or inaccuracies within the utilized cosmological model for generating SNe Ia at varying distances; however, these deviations do not significantly impact overall error estimation.

This study serves as a preliminary investigation of the Mephisto survey. Once Mephisto is fully operational, we can combine it with the LJT to obtain about three spectra during the peak phase of well-observed SNe Ia. Currently, LJT can identify 30–50 SNe Ia every year. With the addition of Mephisto, the sample size of SNe Ia will be significantly increased. The fundamental astrophysics of SNe Ia, including their progenitor scenarios and explosion mechanisms, remains elusive (\citealp{2023RAA....23h2001L}). A larger sample size of SNe Ia can significantly improve our ability to constrain the complex interplay between the progenitor channels, explosion models, and observations.

Although Mephisto may not exhibit a numerical advantage in the quantity of SN Ia discoveries compared to other large-scale survey telescopes, its observational capabilities are formidable when detecting SNe Ia at lower redshifts. Furthermore, the Mephisto project prides itself on its exceptional color calibration precision and unique ability to conduct simultaneous observations across three spectral bands. As a result, these features play a significant role in partially mitigating systematic errors arising from variations in site weather conditions.

We conducted a simulation for Mephisto to investigate the accuracy of cosmological parameter estimation under various observation strategies, with a fixed observation duration of 180 deg$^2$ and 200 days. The optimal observation strategy we have determined is an exposure duration of 130 seconds and a frequency of three days. In our simulation, more than 900 SNe can be detected in one year of observation. Implementing this strategy ensures commendable photometric precision and enables us to detect numerous SNe Ia. Our simulation has considered the influence of the lunar phase in the light curve and assessed the photometric precision of SNe Ia at $z\sim$ 0.1 for various exposure durations. The $\sigma_\mu$ of sources near redshift 0.1 are about 0.1 mag, with distance uncertainty 4.5\%. Compared to the existing sample, the distance measurement accuracy has improved from 7\% to 4.5\%. By incorporating spectroscopic analysis to exclude abnormal SNe Ia, it is possible to achieve a better fit of the Phillips relation coefficients and further enhance distance measurement precision. 
We found that beyond an exposure time of 60 s, there is a diminishing improvement in photometric accuracy with increasing duration. Once Mephisto runs officially for survey purposes, the collaborative observations with LJT will substantially increase the sample size of SNe Ia and facilitate more comprehensive spectroscopic classification, thereby mitigating the scatter induced by the diversities.

Mephisto and LJT can build an extensive SN Ia database, a valuable resource for further analysis. This database will encompass light curves exhibiting diverse shapes and brightness levels, along with spectra captured near the peak of the light curve. This comprehensive dataset will offer a valuable tool for investigating the heterogeneity of SN Ia events and comprehending their utility as cosmological probes. Through meticulous examination of these data, we can derive precise distance measurements, delve into the nature of dark energy, and unveil fresh insights into the origin and evolution of our universe. 

\begin{acknowledgements}
	We thank the anonymous referee for constructive comments
	and suggestions. This work was supported by the National Key R\&D Program of China (2021YFA1600404), the National Natural Science Foundation of China (12173082), and science research grants from the China Manned Space Project (CMS-CSST-2021-A12), the Yunnan Province Foundation (202201AT070069), the Top-notch Young Talents Program of Yunnan Province, the Light of West China Program provided by the Chinese Academy of Sciences, and the International Centre of Supernovae, Yunnan Key Laboratory (202302AN360001). We acknowledge the support of the staff of the LJT and Mephisto. Funding for the LJT has been provided by the CAS and the People’s Government of Yunnan Province. The LJT is jointly operated and administrated by YNAO and the Center for Astronomical Mega-Science, CAS. Mephisto is mainly funded by the “Yunnan University Development Plan for World-Class University” and “Yunnan University Development Plan for World-Class Astronomy Discipline”, and obtained supports from the “Science \& Technology Champion Project” (202005AB160002) and from two “Team Projects” – the “Innovation Team” (202105AE160021) and the “Top Team” (202305AT350002), all funded by the “Yunnan Revitalization Talent Support Program”.

\end{acknowledgements}

\bibliography{ms2023-0407}
\bibliographystyle{raa}

\end{document}